\documentclass[12pt,preprint]{aastex}

\shorttitle{High-Inclination Kuiper Belt Dynamics}
\shortauthors{Kuchner, Brown \& Holman}
\begin{document}
 
\title{Long-Term Dynamics and the Orbital Inclinations of the \\ Classical Kuiper Belt Objects}
\author{Marc J. Kuchner\altaffilmark{1}}
\affil{Harvard-Smithsonian Center for Astrophysics \\ Mail Stop 20, 60 Garden St., Cambridge, MA 02138}
\altaffiltext{1}{Michelson Postdoctoral Fellow}
\email{mkuchner@cfa.harvard.edu}

\author{Michael E. Brown}
\affil{Division of Planetary Sciences, California
Institute of Technology, Pasadena, CA 91125}
\email{mbrown@caltech.edu}

\author{Matthew Holman}
\affil{Harvard-Smithsonian Center for Astrophysics, Mail Stop 18}
\email{mholman@cfa.harvard.edu}

\begin{abstract}
We numerically integrated the orbits of 1458 particles
in the region of the classical Kuiper Belt (41 AU $\leq a \leq$ 47 AU)
to explore the role of dynamical instabilities in sculpting the
inclination distribution of the classical Kuiper Belt Objects (KBOs).  We
find that the selective removal of low-inclination objects by
overlapping secular resonances ($\nu_{17}$ and $\nu_{18}$)
acts to raise the mean inclination of the surviving population of
particles over 4 billion years of interactions with Jupiter, Saturn,
Uranus and Neptune, though these long-term dynamical effects do not
themselves appear to explain the discovery of KBOs
with inclinations near $30^{\circ}$.  Our
integrations also imply that after 3 billion years of interaction
with the massive planets, high inclination KBOs more
efficiently supply Neptune-encountering objects, the likely progenitors
of short-period comets, Centaurs, and scattered KBOs.  The secular resonances
at low inclinations may indirectly cause this effect by weeding out
objects unprotected by mean motion resonances during the first 3 billion 
years.

\end{abstract}
 
 
\section{Introduction}
     
Most of the mass of the Kuiper Belt between 30 and 50 AU appears
to reside in a region outside semimajor axis $a = 40$ AU called
the classical Kuiper Belt \citep{jewi98}.  Early Kuiper Belt surveys
found that Kuiper Belt Objects (KBOs) can have
surprisingly high orbital inclinations \citep{jewi95, jewi96}.
Recent large-scale surveys \citep{jewi98, tjl01} have established that
the classical KBOs---not just the relatively nearby plutinos---frequently have
high orbital inclinations ($i \gtrsim 15^{\circ}$).

The discovery of these high-inclination objects seems
to point to unknown processes in the primordial
solar system.   Several mechanisms for ``pumping-up''
the ancient KBO inclination distribution have been
investigated, such as resonant encounters \citep{malh95,naga00} and
perturbations from passing stars \citep{ida00}, scattered planets
\citep{thom02} and planetesimals \citep{peti99}.  Large KBOs that
may have existed long ago during the epoch of planet formation could also
have stirred the KBO orbits \citep{keny02}.

However, theories of the ancient solar system alone
can not explain the current KBO orbital distribution.
Orbit integrations in the Kuiper Belt
\citep{torb89, torb90, glad90, levi93, holm93}
have shown that the dynamical effects of the massive planets
on the Kuiper Belt have probably removed most of the original KBOs over the
lifetime of the solar system.  Objects removed recently by these
processes probably supply today's population of Centaurs, short-period
comets, and scattered KBOs \citep{fern80, dunc88}.  One might ask
whether selective removal of objects by interactions with the planets has
altered the inclination distribution of the surviving population.

Furthermore, debiased estimates of the KBO
inclination distribution \citep{brow01} suggest that the classical KBOs 
divide into two populations: one dynamically warm, with a typical inclination
of $i \approx 17^{\circ}$, and one dynamically cold, with a typical
inclination $i \approx 2^{\circ}$.  We wonder whether long-term
dynamical interactions helped create the apparent two-component
distribution.

To understand the role of dynamical
stability in shaping the classical KBO inclination distribution and
creating Centaurs, comets, and scattered KBOs, we performed a new
large-scale simulation of the dynamics of KBOs,
integrating the orbits of 1458 test particles in the region of the
classical Kuiper Belt under the influence of gravity from Jupiter,
Saturn, Uranus, and Neptune.  \citet{dunc95} performed the most recent
previous large-scale integration of orbits in the Kuiper Belt.  They
used an initial grid with fine resolution
in semimajor axis, but only sampled a few initial inclinations, and
did not integrate the orbits of high-inclination objects
$i > 1^{\circ}$ for longer than one billion years.  Our work complements
theirs; at the cost of coverage in semimajor axis, we focus
on exploring the consequences of high inclination and
eccentricity over a four-billion-year period.

\section{The Observed Classical KBO Orbits}
\label{sec:observed}

\begin{figure}
\figurenum{1}
\epsscale{1.0}
\plotone{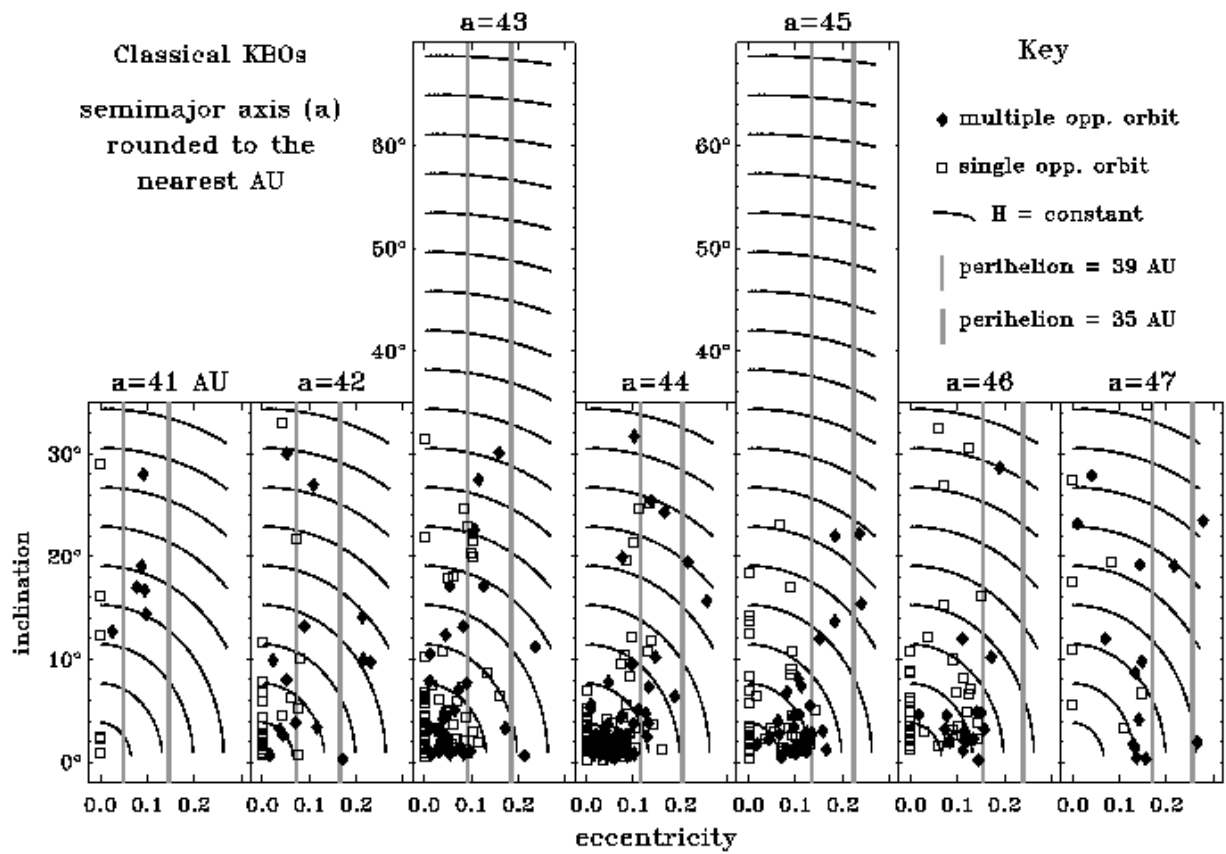}
\caption{Inclinations and eccentricities of the known classical KBOs,
sorted by semimajor axis rounded to the nearest AU.  The vertical lines indicate
perihelion distances of 35 and 39 AU, and the curves show constant values of
$H=\sqrt{G M_{\odot} a (1-e^2)} \, \cos{i}$, the component of angular mementum perpendicular
to the ecliptic plane.}
\label{fig:realkbos}
\end{figure}

Figure~\ref{fig:realkbos} shows the orbital eccentricities and
inclinations with respect to the ecliptic plane of the 410 known
classical KBOs with $40.5 < a < 47.5$ AU sorted into seven panels
by semimajor axis, $a$, rounded to the nearest AU.
We obtained
these data from the Minor Planet Center's website.\footnote{The
Minor Planet Center's website is available at
\url{http://cfa-www.harvard.edu/iau/mpc.html}.}
Filled symbols indicate orbits for KBOs observed at multiple oppositions.
The single-opposition orbits, shown as empty squares, generally
have accurate inclinations, but poorly measured eccentricities; often
the Minor Planet Center assumes zero eccentricity for single-opposition
orbits.

These classical KBOs have a mean inclination of $6.6^{\circ}$.  However,
the data reflect a strong observational bias against detecting
high-inclination KBOs; high inclination objects spend less time near
the ecliptic plane than low-inclination objects.  
The figure also reflects severe observational biases against detecting distant
KBOs.  The brightness of a KBO in reflected light decreases roughly
as $1/r^4$, where $r$ is the KBO's heliocentric distance.  This
distance bias propagates into correlated biases in semimajor axis, eccentricity,
and mean anomaly, since most KBOs are discovered near perihelion.
See \citet{truj01} and \citet{glad01} for recent discussions of the
radial distribution of KBOs.

Plotted on Figure~\ref{fig:realkbos} is a family of curves corresponding to
constant values of $H=\sqrt{G M_{\odot} a (1-e^2)} \, \cos{i}$, the
component of the orbital angular momentum (per particle mass) in the
direction perpendicular to the ecliptic plane.  $H$ is largest in the
lower left-hand corner of each plot ($i = e =0$).
In a system where all the planets occupy
circular orbits in the same plane, the orbit of a test particle
conserves the component of its angular momentum perpendicular to that plane
\citep{koza62,thom96}.  We would expect the test particles in our simulation
to conserve $H$ and only evolve parallel to these curves
in the absence of semimajor axis variations if all the
perturbing planets had $i=e=0$ with respect to
the ecliptic plane.

\begin{figure}
\figurenum{2}
\epsscale{1.0}
\plotone{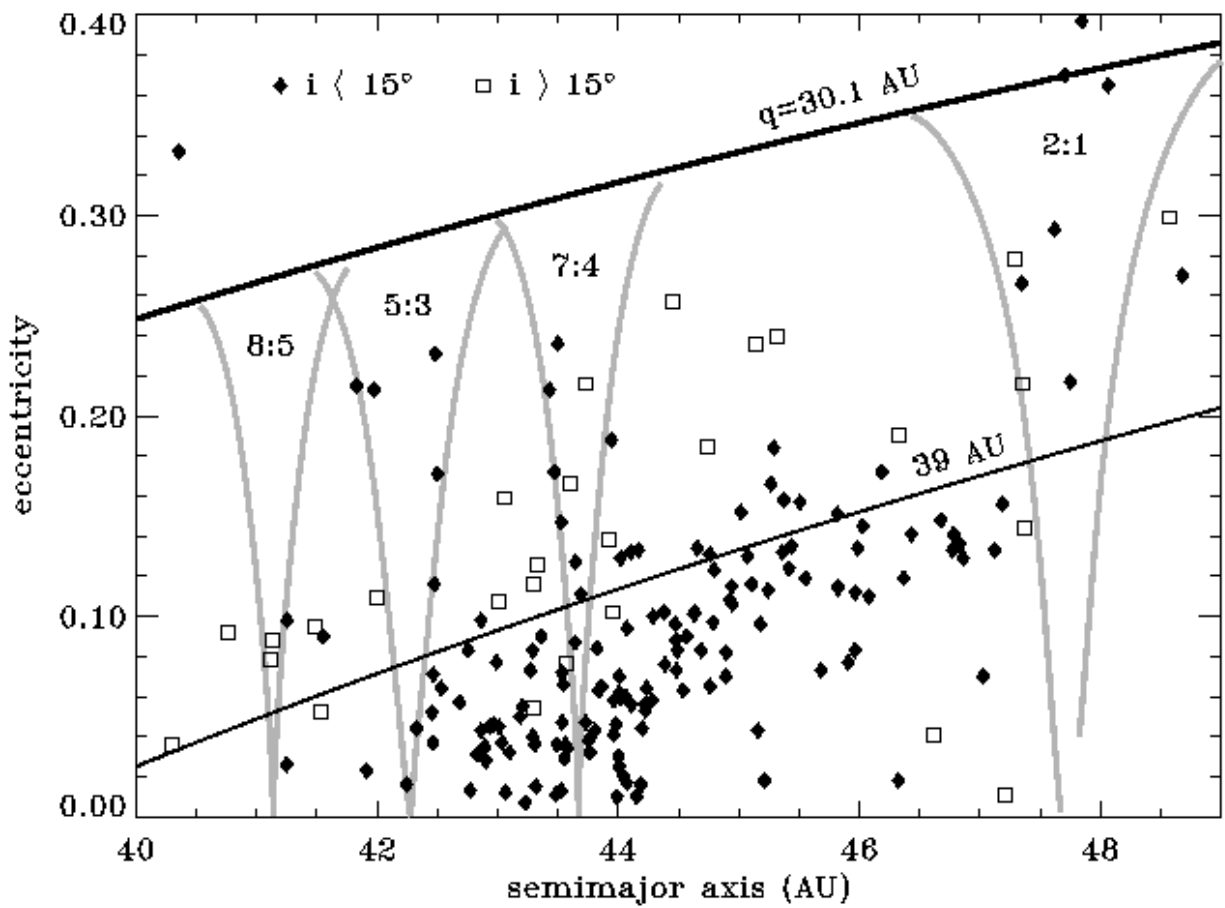}
\caption{Eccentricity vs. semimajor axis for the orbits of the known classical
KBOs observed at more than one opposition.  High inclination orbits
($i > 15^{\circ}$) are squares; low-inclination ($i < 15^{\circ}$) orbits are
filled diamonds.  Grey curves indicate numerically computed boundaries
of mean motion resonances (MMRs) with Neptune, and black curves show where
the perihelion distance, $q$, is 30.1 AU, i.e., the KBOs become
Neptune crossing, (thick curve) and where $q=39$~AU (thin curve).}
\label{fig:realae}
\end{figure}

Figure~\ref{fig:realae} shows the eccentricities and semimajor axes
of the known KBOs observed at more than one opposition.
Two curves on this figure show the eccentricities corresponding to 
perihelion distances of $q=30.1$ AU (the Neptune crossing line)
and $q=39$ AU.  Also, each panel in Figure~\ref{fig:realkbos} has 
vertical grey lines indicating $q=35$ AU and $q=39$ AU.
The dearth of objects with $q$ much greater than 39 AU in
Figure~\ref{fig:realae} shows the bias towards detecting objects with small
perihelia.

Forty-seven of the classical KBOs with observations at multiple
oppositions have orbits with perihelia less than 39 AU.  
These 47 objects have a mean inclination of $13^{\circ}$, much
higher than the mean inclination of the classical KBOs as a
whole ($6.6^{\circ}$).  Only 11 of these 47 have orbits with
perihelion $q < 35$ AU.  \citet{dunc95} found that particles
with $q < 35$ AU generally did not survive one billion years of
interaction with the giant planets.


Grey curves in Figure~\ref{fig:realae}
indicate the boundaries of the 8:5, 5:3, 7:4 and 2:1 mean motion
resonances (MMRs) with Neptune, using the planar circular
restricted three-body problem as a model for the motion of
small bodies in the field of the Sun and Neptune.  We numerically
computed these boundaries following the technique developed
by \citet{ferr89} to avoid relying on analytic expansions of
the disturbing function which invoke low eccentricities.
\citet{morb95} first applied this approach to MMRs
in the Kuiper belt.  The resonance widths calculated this way contain
a chaotic zone along the boundary at high eccentricities, while
the widths calculated by \citet{malh96} exclude this region.  This
chaos may affect the long-term dynamics and stability of KBOs along
the resonance boundaries.

\section{Orbit Integrations}
 
We integrated the orbits of 1458 non-interacting particles for 4 billion
years under the gravitational influences of Jupiter, Saturn,  Uranus,
Neptune and the Sun using the MVS symplectic integrator from
the SWIFT package by \citet{levi94} based on an algorithm by \citet{wisd91}.
The initial arguments of pericenter, mean longitudes, and
ascending nodes were chosen randomly from a uniform distribution on the interval
[0,2$\pi$).  We used orbital elements referred to the ecliptic plane,
rather than the invariable plane, so they can easily be compared with
observations.   We used orbital data for the planets
from \citet{cohe73} and time steps of one year.
The integration required donated time on 20 different Sun workstations
at the Caltech Astronomy Department and at the Harvard-Smithsonian
Center for Astrophysics.  To minimize network traffic, we recorded
the orbital elements of each particle only once every 10 million years.

To check that our choice of a one year step size was small enough,
we integrated the orbits of 100 objects for a billion years in the
vicinity of a subtle high-inclination feature
(a=42, e=0, i=[60,60.02,60.04,\ldots,61.98]; see Section~\ref{sec:kozai}),
using a step size of 1 year.  Then we repeated the integration using
the same initial conditions and a step size of 1/4 of a year.  Naturally,
the final orbits differed due to the non-linearity of the dynamics
and the long integration times.  But the final distributions of the
orbits in eccentricity and inclination space were indistinguishable.

Our initial population of objects had semimajor
axes of [{41, 42, 43,\ldots, 47}] AU, eccentricities of [{0, 1/30,
2/30,\ldots, 8/30}] and inclinations of
[{0, 1/30, 2/30,\ldots, 17/30}] radians ([{0.0000, 1.9099, 3.8197,\ldots,
32.4676}] degrees).   
We refer to this grid of 1134 particles, which evenly fills a rectangular
prism in ($a$, $e$, $i$) space, as our first batch.  We also performed some
exploratory integrations with higher initial inclinations; at 43 and 45 AU, we started
particles with inclinations up to 1.1666 radians (66.8451 degrees).
We removed particles from the integration when they encountered Neptune's
Hill sphere.  Six hundred and twenty-six particles from the first batch
survived the entire integration.  

\subsection{Semimajor Axis Evolution}
\label{sec:semimajor}

Though the particles began with integral values of semimajor axis,
after 10 million years of integration only a shadow of the
discrete initial conditions remained.  If we combine all the data
in the first batch, we find that
by $t=10$ million years, the particles have spread out in semimajor axis
into a distribution with half-width at half maximum of 0.4 AU.
At $t=4$ billion years, the core of the distribution has not changed, but
the particles in the wings are gone.  Only 0.1 \% of the particles that
survived for 4 billion years were transported farther than 0.7 AU
from their initial semimajor axes.  We infer that the observed classical
KBOs have not strayed far in semimajor axis since the planets
settled into their current configuration---unless another effect
besides long term interactions with the planets disturbed their orbits.

\begin{figure}
\figurenum{3}
\epsscale{1.0}
\plotone{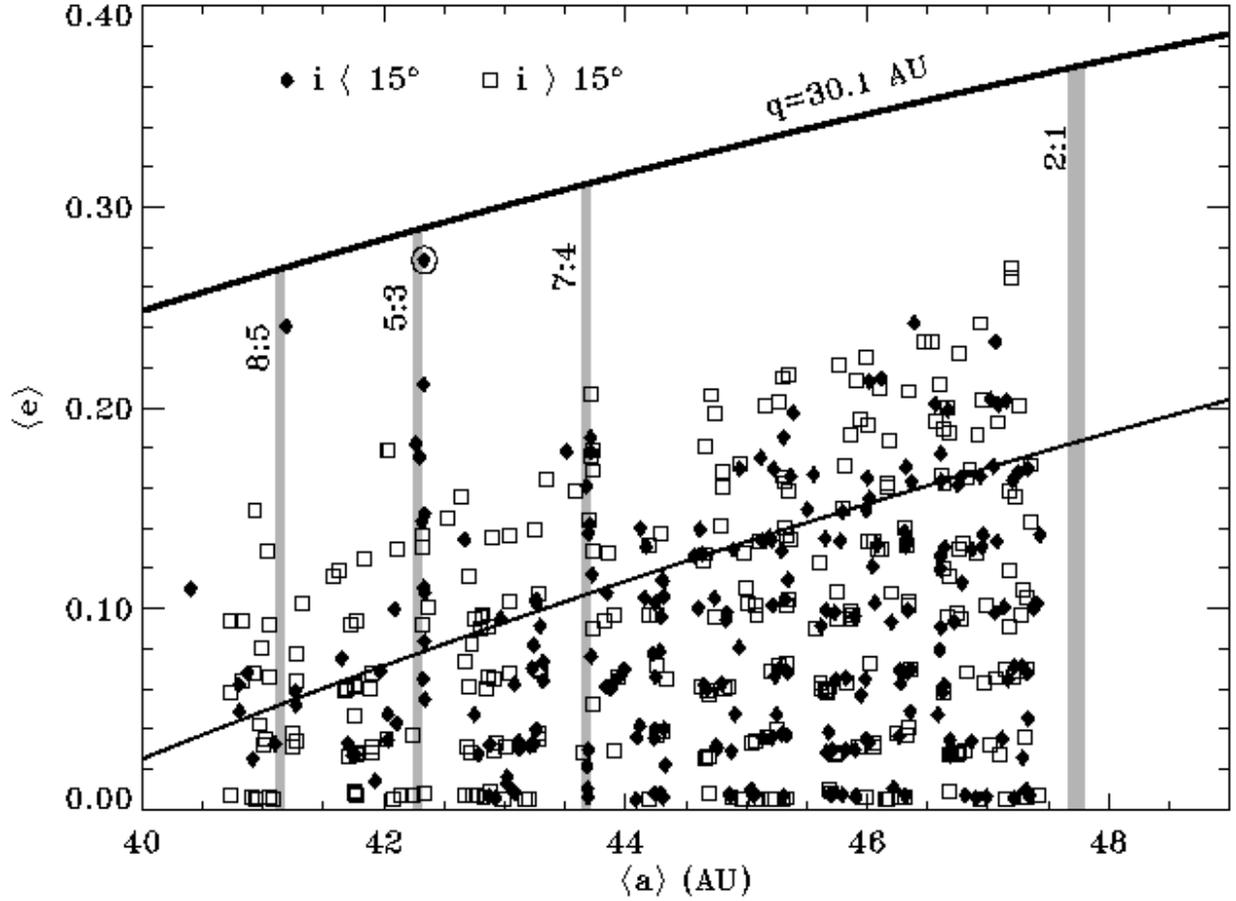}
\caption{Mean eccentricity vs. mean semimajor axis for the particles
in the first batch of our integration that survived for 4 billion years,
averaged over the period $t=3$ billion to $t=4$ billion years.
Vertical grey lines indicate the nominal locations of mean motion resonances
(MMRs) with Neptune, and black curves show where $q=30.1$~AU 
(thick curve) and where $q=39$~AU (thin curve).}
\label{fig:simulatedae}
\end{figure}

Figure~\ref{fig:simulatedae} shows the eccentricities
and semimajor axes of the particles that survived for 4 billion years.
The eccentricities and semimajor axes have been averaged over the period
3-4 billion years (100 data dumps).  Solid diamonds show the averaged
orbits of low inclination test particles ($i < 15^{\circ}$); empty squares
show the averaged orbits of high-inclination particles
($15^{\circ} \le i < 30^{\circ}$).  In the lower right, particles
at low $e$ and high $a$ remain clustered, reflecting the initial
conditions.   This zone appears to be stable for the lifetime of
the solar system, however few KBOs
are known with high $a$ and low $e$ orbits, since the perihelia of these
orbits are far from the sun.  To the left of the figure, at $a < 44$ AU,
an inclination-dependent instability appears to have removed many
of the low-inclination particles, particularly at perihelia $q <39$ AU.

The grey vertical lines indicate the nominal positions of
the 8:5, 5:3, 7:4 and 2:1 MMRs.  Since the
plotted orbits are averaged over the resonant librations, the libration
widths of the resonances do not appear.  
Figures~\ref{fig:realae} and \ref{fig:simulatedae} show that
objects with small perihelion distances, $q$, often
inhabit MMRs, which can protect them from close encounters with
Neptune.  The MMRs at $a < 44$ AU appear to serve double
protective duty;  they protect low $q$ objects from close encounters,
and protect low inclination objects from ejection by the
instability that removes low inclination particles
(see Sections~\ref{sec:secularresonances} and \ref{sec:comet}).

\subsection{Eccentricity and Inclination Evolution}
\label{sec:eccentricity}

\begin{figure}
\figurenum{4}
\epsscale{1.0}
\plotone{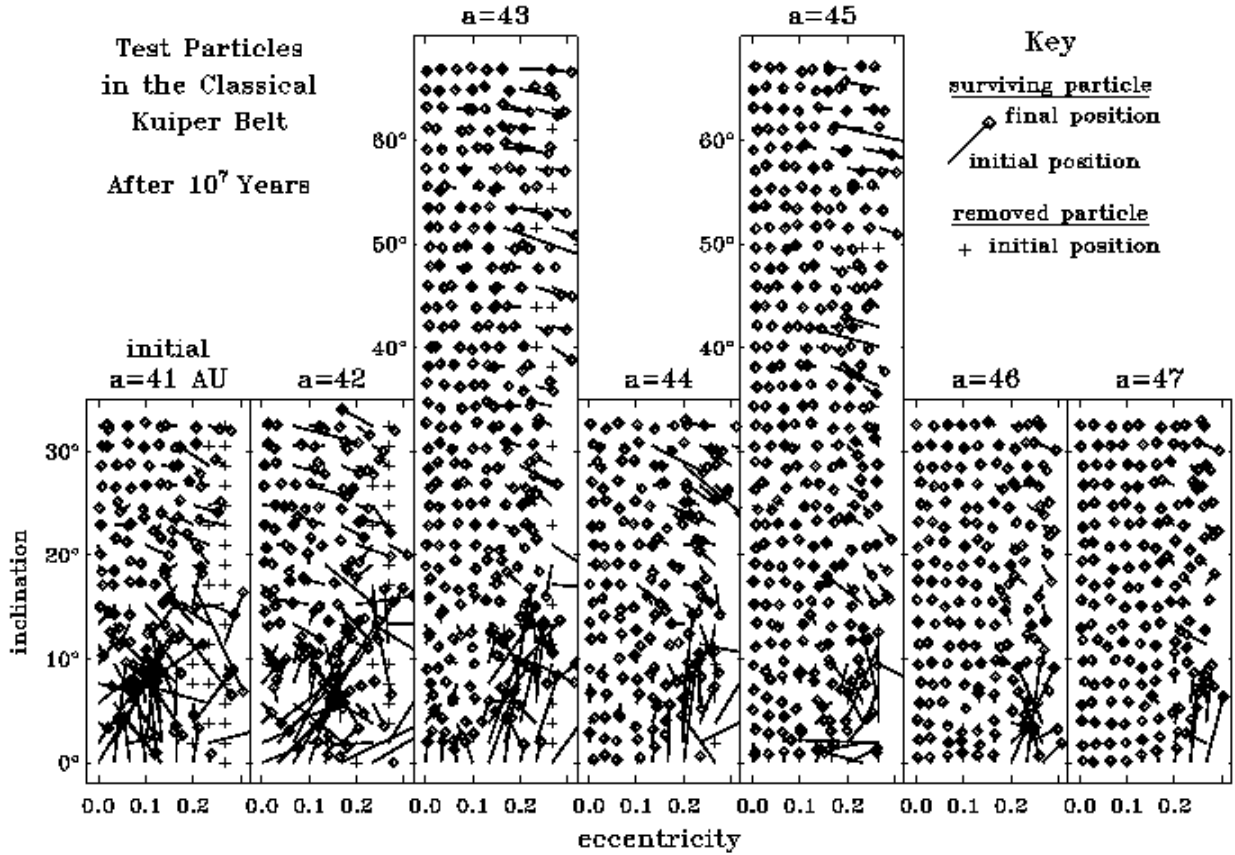}
\caption{After 10 million years, the orbits of our test particles have
evolved to the locations indicated by the diamonds.  Short lines connect
each diamond to the initial orbit.  Crosses show the initial locations of
particles that have been removed.  The particles are sorted into seven
panels by their initial semimajor axes.}
\label{fig:tenmillion}
\end{figure}

Figure~\ref{fig:tenmillion} shows the orbits of the test particles 
after ten million years of integration.  The particles are sorted into
seven panels by their initial semimajor axes.  Diamonds indicate the
orbital eccentricity and inclination of the particles that survived
the integration for ten million years.  Thin lines connect the diamonds
to the initial orbits.  Crosses indicate the initial positions of
particles that suffered close encounters with Neptune.

Even after this relatively short period, many of the particles at
low inclinations $i < 15^{\circ}$ have been substantially perturbed,
their inclinations raised.  Particles with inclinations
$i \gtrsim 20^{\circ}$ have generally evolved along lines of $H=$constant, shown
in Figure~\ref{fig:realkbos}.  But so far, only a few low-inclination
particles have been removed.

\begin{figure}
\figurenum{5}
\epsscale{1.0}
\plotone{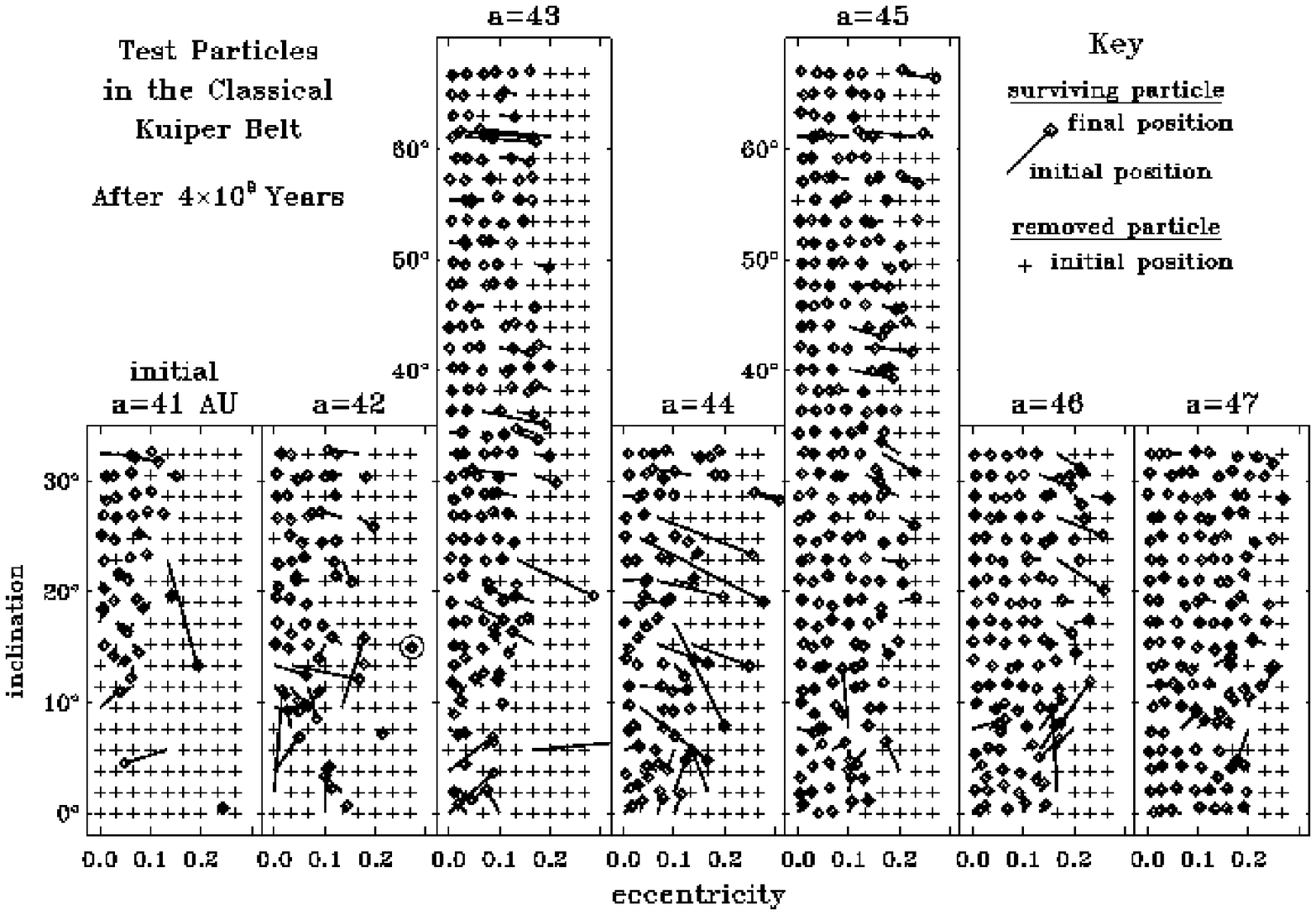}
\caption{After four billion years, most of the particles whose orbits 
dramatically evolved in the first ten million years (the diamonds with the long tails in
Figure~\ref{fig:tenmillion}) are gone.   The $\nu_{8}$, $\nu_{18}$, and $\nu_{17}$ secular resonances
have removed some low-inclination particles at initial semimajor axes $a=41$ and $42$ AU.
Evolution at high inclinations is mostly parallel to the curves of
$H=$constant shown in Figure~\ref{fig:realkbos}.}
\label{fig:fourbillion}
\end{figure}

Figure~\ref{fig:fourbillion} shows the state of affairs after four billion years,
summarizing the inclination and eccentricity evolution of the particles in
the same manner as Figure~\ref{fig:tenmillion}.  This figure should be compared to
the current distribution of KBO orbits (Figure~\ref{fig:realkbos}).
By this time, most of the
low inclination particles whose orbits were greatly perturbed
(the diamonds with the long lines in Figure~\ref{fig:tenmillion}) have now
been ejected.
By four billion years, nearly all of the objects with initial perihelion
$q< 35$ AU have been removed.  This cutoff is not even: several objects
have been transported into the region $q < 35$ AU, and at low
inclinations $i \lesssim 10^{\circ}$, the border of the unstable
region is closer to $q = 38$ AU (see Figure~\ref{fig:realkbos}).
However, Figure~\ref{fig:fourbillion} shows a systematic removal of
high-eccentricity objects, which vividly demonstrates
that creating the high-inclination KBOs does not require a
special mechanism that only increases inclinations and not eccentricities.

When we discuss the relevance of our simulations to the KBO
inclination distribution, we find it convenient to define two zones in the
classical Kuiper Belt, one at $40.5$ AU $< a < 44$ AU, the other
at $44$ AU $< a < 47.5$ AU.  This division cuts the distribution of
known KBOs roughly in half; of the classical KBOs plotted in
Figure~\ref{fig:realkbos}, 213 fall in the first group (84 with multiple
opposition orbits) and 197 fall the second group (92 with
multiple opposition orbits).  The debiased estimate of the radial
distribution of KBOs by \cite{truj01} is also roughly symmetric
about 44 AU.  As Figure~\ref{fig:simulatedae} shows, 
most of the inclination dependent structure after 4 billion years
is located in the first zone, at $a < 44$ AU.

\begin{figure}
\figurenum{6}
\epsscale{0.8}
\plotone{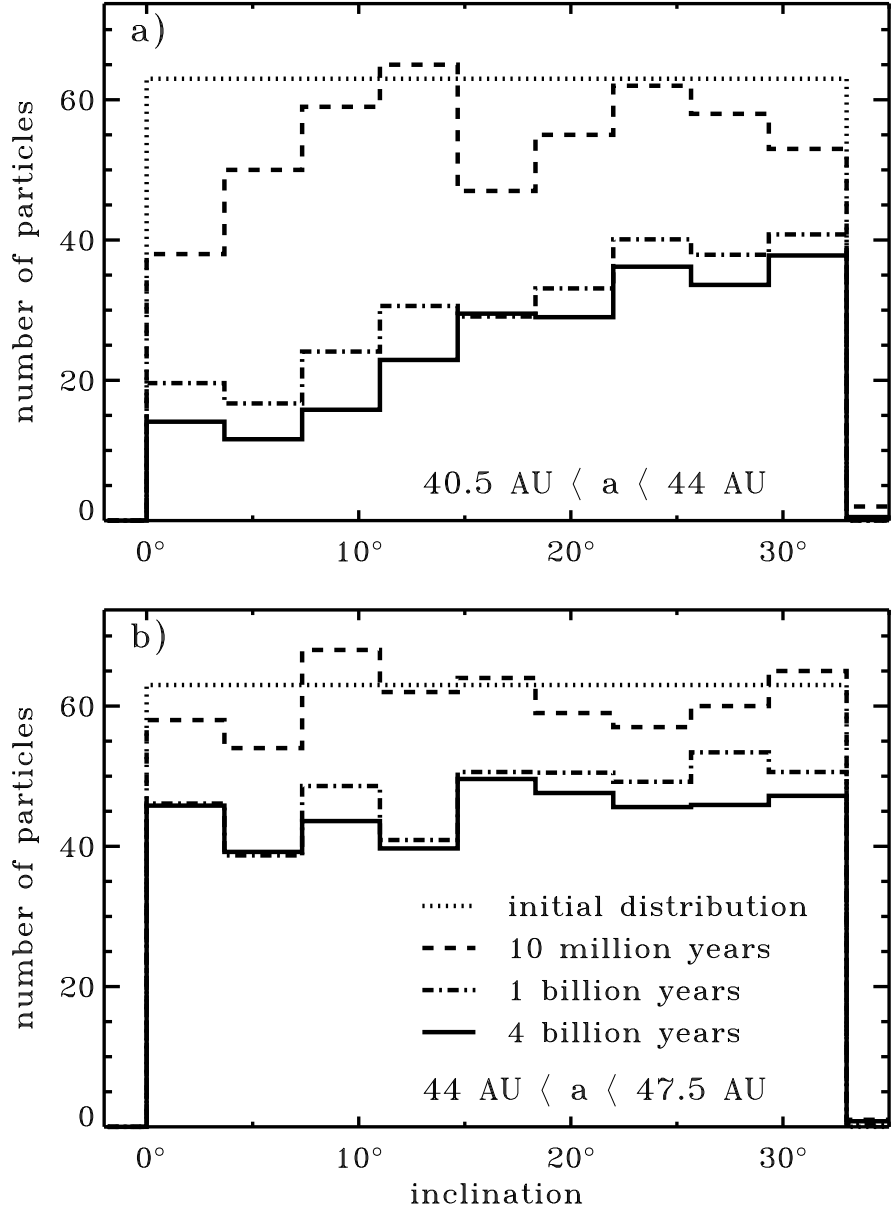}
\caption{Histograms of the orbital inclinations of the surviving
particles from our first batch of integrations ($i < 33^{\circ}$) after
0 years (dotted line), 10 million years (dashed line), 1 billion years
(dash-dot line), and 4 billion years (solid line).
a) Only particles with $40.5$ AU $< a < 44$ AU
b) Only particles with $44$ AU $< a < 47.5$ AU.} 
\label{fig:inchistogram}
\end{figure}

Figure~\ref{fig:inchistogram} summarizes the inclination distributions of
the two zones, using the first batch of particles ($ i < 33^{\circ}$)
after 10 million, 1 billion, and 4 billion years.
The semimajor axis limits are applied when the distributions are
evaluated.  The 1 billion and 4 billion year distributions were
averaged over 100 million years (10 data dumps) to cover periodic
secular motions and provide better statistics.

After 10 million years (the dashed line), only 9\%
of the particles have been ejected, but many of the low inclination objects
have moved to higher inclination orbits.  After 4 billion years,
almost half of the particles (46\%) are gone, and
the region $40.5$ AU $< a < 44$ AU has been hit the hardest, especially at
low inclinations.
An approximation to the final distribution at 4 billion years
(the solid curves in Figure~\ref{fig:inchistogram}) is
\begin{equation}
n(i) \approx
\left\{ \begin{array}{ll}
0.20 + 0.0017 (i - 4^{\circ})^{2} & 40.5\ \rm{AU} < a < 44\ \rm{AU},\ i < 17.5^{\circ} \\
                             0.51 & 40.5\ \rm{AU} < a < 44\ \rm{AU},\ i \ge 17.5^{\circ} \\
                             0.7  & 44\ \rm{AU} < a < 47.5\ \rm{AU}
\end{array}
\right.
\label{eq:approximation}
\end{equation}

\section{The Causes of the Instabilities}

As far as we know, all dynamical instability is caused by the overlap of
resonances.  We performed some additional integrations to help understand
the dynamics of the unstable regions in the classical Kuiper Belt
in this context.

\subsection{Secular Resonances at Low Inclinations}
\label{sec:secularresonances}

The only notable secular resonances from 41--47 AU are the perihelion and
node resonances with Neptune ($\nu_{8}$ and $\nu_{18}$) and the node resonance with Uranus ($\nu_{17}$),
all of which are nominally located at $a < 41.6$ AU \citep{knez91}.
Figure~3 of \citet{dunc95} shows that these resonances deplete objects at
low inclinations at $a=$41--42 AU.  However, Figure~4 of \citet{holm93}
shows that these resonances affect the inclinations of particles
as distant as $a=50$ AU.  Our Figure~\ref{fig:tenmillion} 
shows in three dimensions a region of stirring in inclination and
eccentricity that penetrates all of the 7 panels at low inclinations
and high eccentricities.  This region appears to be the broad wing
of the zone of influence of the overlapped secular resonances centered
at 41--42 AU, shown in Figure~4 of \citet{holm93}.

To test this identification, we integrated the orbits of 72 particles with
initial $a=43$ AU, $e=0.0$,
$\Omega=[0^{\circ},54^{\circ},126^{\circ},198^{\circ},270^{\circ},342^{\circ}]$,
$i=[0, 1/50, 2/50, \ldots , 11/50]$ radians ($0^{\circ}$--$12.6051^{\circ}$).  Ten of
these particles were removed before 4 billion years, all from initial inclinations
$\le 6/50$ radians ($6.875^{\circ}$).  We plotted pseudo-surfaces of section
for all of these particles in various coordinate systems to look for librational
motion indicating the presence of a resonance.  

\begin{figure}
\figurenum{7}
\epsscale{1.0}
\plotone{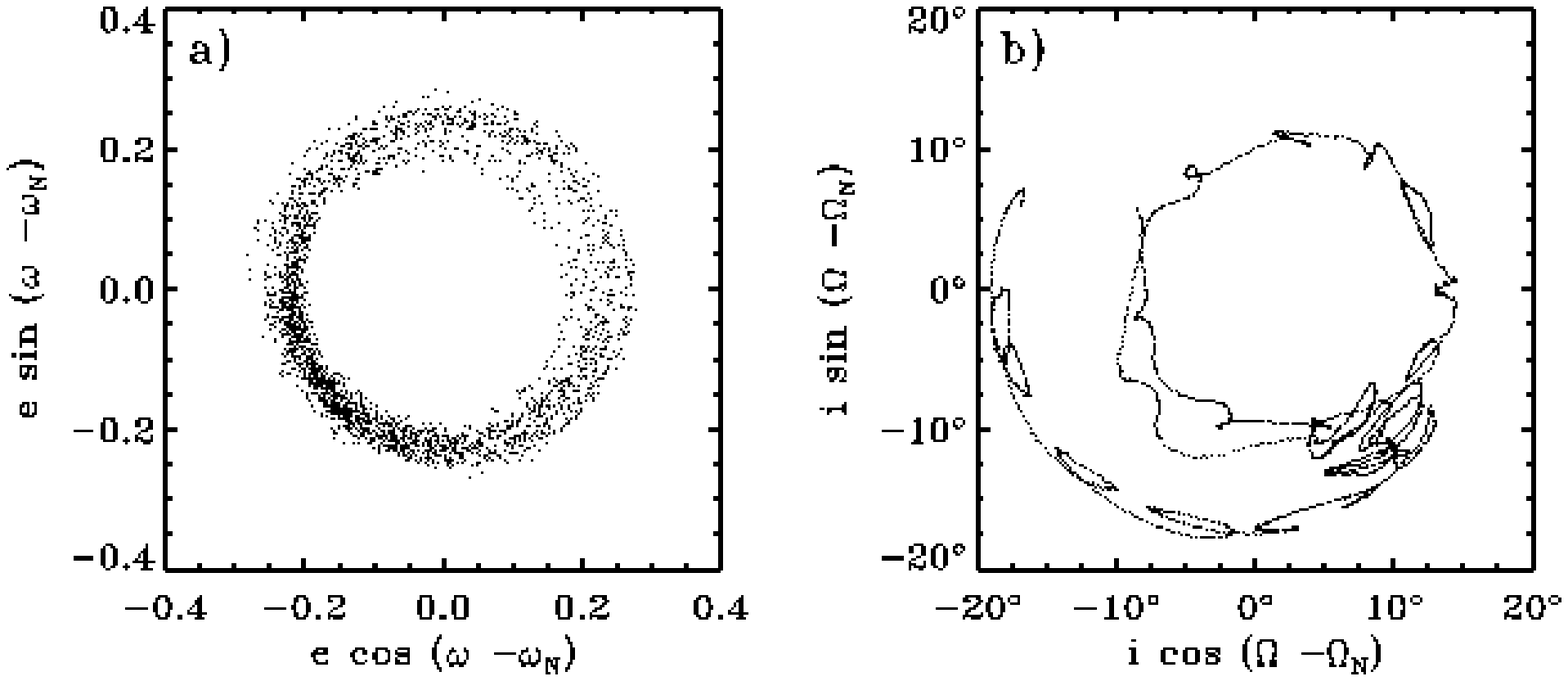}
\caption{Pseudo-surfaces of section for a typical ejected
low-inclination particle with initial semimajor axis $a=43$ AU,
eccentricity $e=0$.
a) A polar plot of eccentricity and $\omega - \omega_{N}$
b) A polar plot of inclination and $\Omega - \Omega_{N}$.}
\label{fig:lowi43}
\end{figure}

All of the particles in this group that were ejected
shared a common behavior.  Figure~\ref{fig:lowi43} shows pseudo-surfaces of section
for a typical ejected particle in two coordinate systems:
$(e \cos{(\omega -\omega_{N})}, e \sin{(\omega -\omega_{N})})$
and $(i \cos{(\Omega -\Omega_{N})}, i \sin{(\Omega -\Omega_{N})})$, where
$\omega$ refers to the argument of pericenter, $\Omega$ refers to
the argument of ascending node, and the subscript N refers to Neptune.
The angle $\omega -\omega_{N}$ circulates,
while the orbit undergoes complex motion in the inclination and
$\Omega -\Omega_{N}$ coordinates, alternating between libration and circulation,
while the inclination slowly grows.
Finding this common behavior 
indicates that competing nodal resonances ($\nu_{17}$ and $\nu_{18}$)
dominate the long-term dynamics of these particles.
These competing resonances appear to cause the broad swath of
destabilization at low inclinations and high
eccentricities apparent in Figures~\ref{fig:simulatedae}, \ref{fig:tenmillion},
\ref{fig:fourbillion}, and \ref{fig:inchistogram}.

\subsection{The Kozai Resonance}
\label{sec:kozai}

The Kozai resonance is the tendency of high inclination orbits in the
three-body problem to undergo secular libration of the argument of
pericenter, $\omega$, coupled with oscillations of
$i$ and $e$, which conserve $H$.  Recall that we measure an orbit's
argument of perihelion from the ascending node, so in the solar system,
this motion corresponds to an oscillation of $\omega$ about the
invariable plane.  \citet{koza62} made the approximation
that the planets all orbit in the same plane, and calculated that
in the solar system, these oscillations set in at
inclinations of $i > 32^{\circ}$ for $a=3$ AU.
\citet{thom96} calculated that under the same approximation,
no region of the Kuiper Belt should be subject to the Kozai resonance.

\begin{figure}
\figurenum{8}
\epsscale{1.0}
\plotone{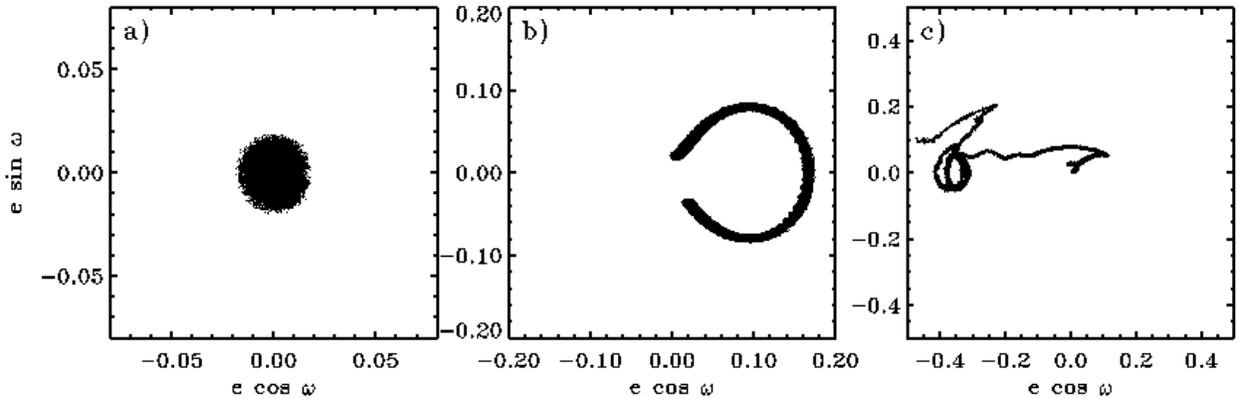}
\caption{Pseudo-surfaces of section for three particles with initial
inclination $i=61^{\circ}$, semimajor axis $a=43$ AU, and eccentricity $e=0$
illustrating the three kind of behaviors we found near $i=61^{\circ}$.
a) Stable chaotic motion confined to low eccentricities.  b) Stable
chaotic motion combined with slow libration around $\omega=90^{\circ}$ or
$\omega=270^{\circ}$.  c) A particle which is ejected under the influence of
the Kozai resonance.  Note the three different scales.}
\label{fig:kozai}
\end{figure}

But Figure~\ref{fig:fourbillion} shows that the surviving particles
around $i = 61^{\circ}$ show more eccentricity evolution
than other high inclination particles.  The strong inclination
dependence of this effect and the way the motion appears to conserve $H$
suggest that it is a Kozai phenomenon.
We integrated the orbits of 24 particles with initial
$a=43$~AU, $e=\Omega=0$ at initial $\omega=[0, 54^{\circ}, 126^{\circ},
198^{\circ}, 270^{\circ}$,$342^{\circ}]$
and initial $i=[60^{\circ}, 61^{\circ}, 62^{\circ}, 63^{\circ}]$
and examined the evolution of their orbits to look for librations,
using the coordinates $h=e \cos{\omega}$ and $k=e\sin{\omega}$.
Figure~\ref{fig:kozai} shows three representative pseudo-surfaces of
section that illustrate the three types of behavior 
we found in this region.  Figures~\ref{fig:kozai}a and \ref{fig:kozai}b
show the evolution of particles that survived the integration, while
Figure~\ref{fig:kozai}c shows the evolution of one of the two
particles (initial $\omega=0, i=60^{\circ}$ and $\omega=0, i=61^{\circ}$)
that did not survive.  The points in Figure~\ref{fig:kozai}a fill a roughly
circular region, indicating chaotic motion confined to low
eccentricities \citep{heno83}.   In Figures~\ref{fig:kozai}b and c, the
motion appears to be a superposition of this chaotic motion with libration about
$\omega=0^{\circ}$ or $\omega=180^{\circ}$, indicated by the loops.
The librational motion indicates the influence of a Kozai resonance.

Occasionally, a particle at low inclinations in a MMR can also undergo
Kozai libration \citep{morb97}.  When we plotted pseudo-surfaces of
section for all of the 1458 particles in the big integration, we found one
particle, with initial $a=42$ AU, $i=15.278^{\circ}$, $e=0.266666$, that
went into libration about $\omega=90^{\circ}$.  The orbit of this particle,
circled in Figures~\ref{fig:simulatedae} and \ref{fig:fourbillion},
is inside Neptune's 5:3 MMR.   The Kozai libration may help the MMR protect
the particle from doom at the hands
of the overlapping secular resonances in this region of dynamical space,
since the particle survives for the entire 4 billion years of the integration.

\section{The Observed KBO Inclination Distribution}
\label{sec:ancient}

\begin{figure}
\figurenum{9}
\epsscale{0.8}
\plotone{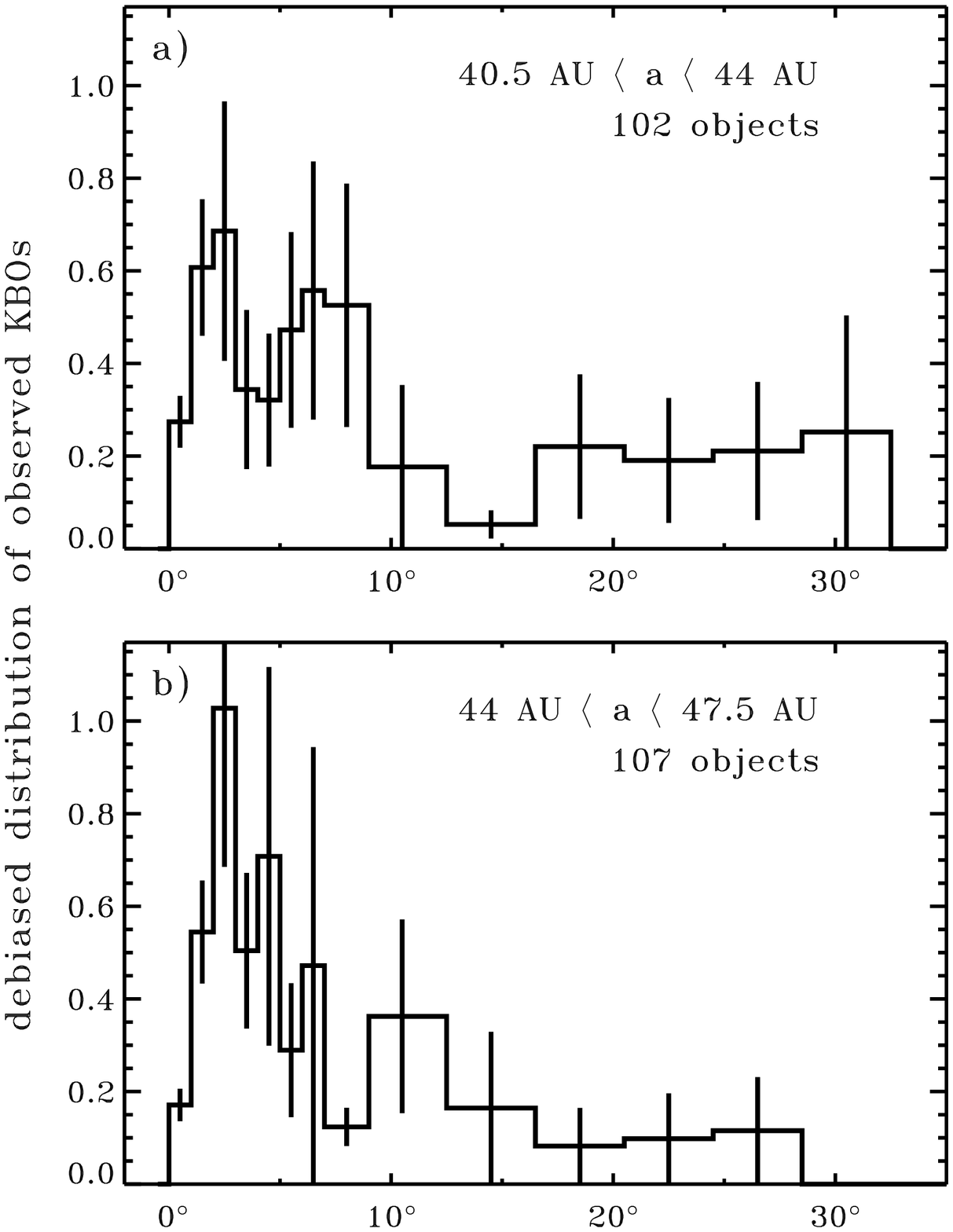}
\caption{Debiased inclination distributions of the classical KBOs
discovered at ecliptic latitude $< 1^{\circ}$.  a)  Objects with semimajor axis
$40.5 < a < 44$.  b) Objects with semimajor axis $44 < a < 47.5$.
Distribution a reflects much more severe sculpting due to
interaction with the massive planets.}
\label{fig:observedinc}
\end{figure}

As we mentioned in Section~\ref{sec:observed}, high-inclination KBOs
spend less time near the ecliptic plane than low inclination KBOs, so 
they evade ecliptic plane surveys.  \citet{brow01} realized
that this observational bias could easily be corrected for the subset of
low-eccentricity KBOs discovered near the ecliptic plane; for this
subset, high inclination objects will be underrepresented by the
simple factor of $\sin{i}$, assuming no correlation between longitude
and inclination.  The Kozai phenomenon could produce such a correlation,
but while Kozai libration may be common among plutinos, our integrations
in the classical Kuiper Belt produced only one Kozai librator at
inclinations $< 50^{\circ}$.  We used the
$\sin{i}$ debiasing method to calculate separate debiased
inclination distributions for the inner part of the classical Kuiper Belt
($40.5 < a < 44$ AU) and the outer part of the classical Kuiper Belt
($44 < a < 47.5$ AU) considering only objects discovered at 
ecliptic latitude $< 1^{\circ}$, 209 objects in total. 
Figure~\ref{fig:observedinc} shows these two distributions, with error
bars indicating the Poisson noise due to the limited number of objects
per bin.

As Figure~\ref{fig:inchistogram} shows, 4 billion years of interaction with
the massive planets affects the inclination distribution of particles in 
the outer half of the classical Kuiper Belt little, but it
could substantially alter the inclination distribution of particles
with $a < 44$ AU.  Perhaps the inclination distribution of the inner
half of the classical Kuiper Belt (Figure~\ref{fig:observedinc}a) once
resembled the inclination distribution of the outer
half of the classical Kuiper Belt (Figure~\ref{fig:observedinc}b).  We have
too few objects to make a detailed comparison, but the
distribution in Figure~\ref{fig:observedinc}b clearly has
relatively more low-inclination objects than the distribution in
Figure~\ref{fig:observedinc}a.

Figure~\ref{fig:observedinc}a shows a modest peak at
$i \approx 8^{\circ}$ which piques our curiosity.
After only 10 million years of interaction with the massive planets, we
find that a substantial fraction of test particles at low initial
inclinations $i \lesssim 5^{\circ}$ have been promoted to orbits with
higher inclinations $i \sim 10^{\circ}$ (see Figure~\ref{fig:tenmillion}
and Figure~\ref{fig:inchistogram}).  This promotion generally fortells
an eventual close encounter with Neptune.  Perhaps the peak at
$i \approx 8^{\circ}$ in Figure~\ref{fig:observedinc}a represents
KBOs recently delivered to $i \sim 0^{\circ}$ by some self-interaction
of the KBO disk and promoted to $i \approx 8^{\circ}$ on their way toward
being hurled into Neptune's Hill sphere by the overlapped $\nu_{17}$
and $\nu_{18}$ secular resonances.

\section{Comet and Scattered KBO Formation}
\label{sec:comet}

The classical Kuiper Belt may be the primary source of short-period comets,
Centaurs, and scattered Kuiper Belt Objects.  \citet{levi97} found that objects 
from the Kuiper Belt that wander into the solar system after a close
encounter with Neptune have a median dynamical lifetime of only
$4.5 \times 10^{7}$ years, so perhaps a substantial fraction
of short period comets and scattered KBOs were classical KBOs less than a
billion years ago.  To explore the
origin of short-period comets, Centaurs, and scattered KBOs, we examined the
particles in our integration which were ejected last from the
classical Kuiper Belt.

\begin{figure}
\figurenum{10}
\epsscale{1.0}
\plotone{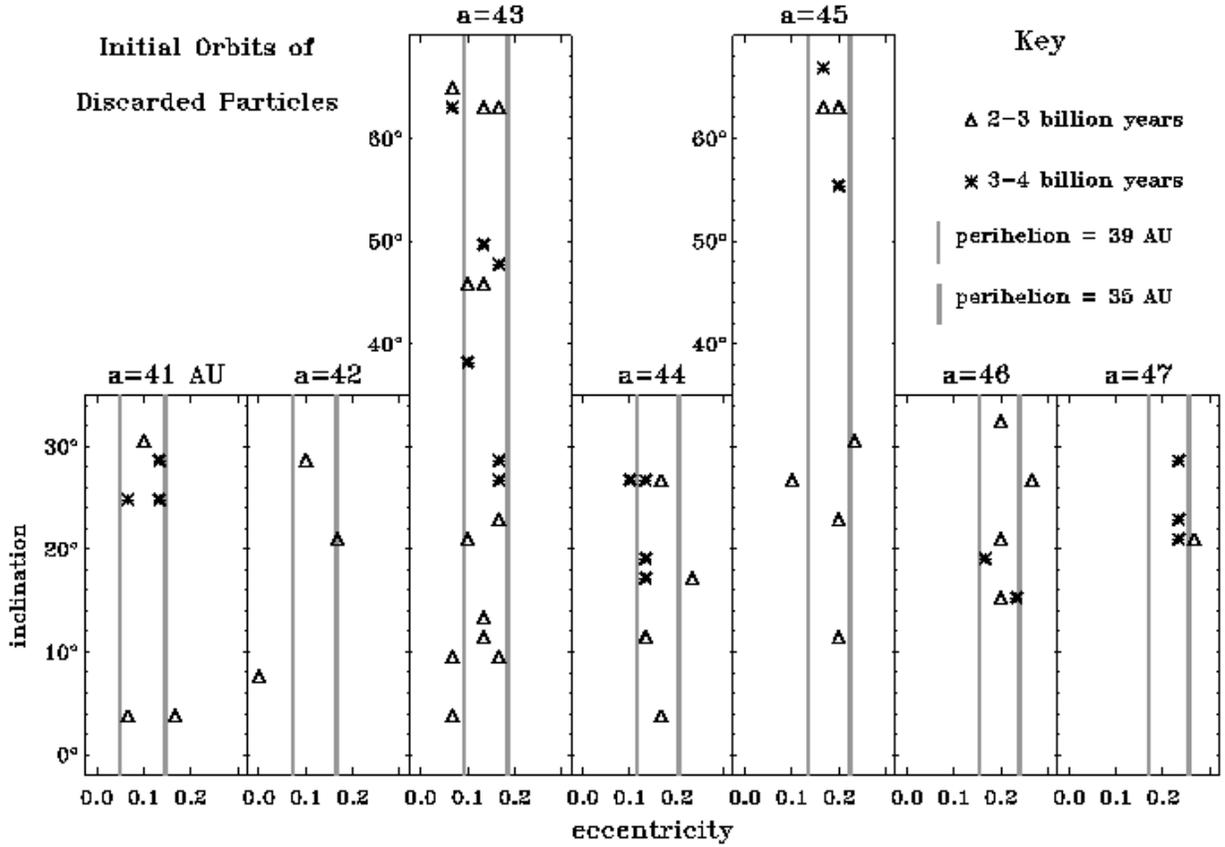}
\caption{Initial orbits of particles that are removed late in the integrations
when they enter Neptune's Hill sphere.  Most of these particles originate on orbits
with perihelia between 35 and 39 AU  (the vertical grey lines).   Particles removed between $t=2$ billion
and $t=3$ billion years (the triangles) come from all inclinations.  Particles removed
between $t=3$ billion years and $t=4$ billion years originated only at high
inclinations ($i > 15^{\circ}$).}
\label{fig:dead}
\end{figure}

Figure~\ref{fig:dead} shows the initial semimajor axes, eccentricities
and inclinations of the particles in our integrations that
encountered Neptune between $t=2$ billion and $t=4$ billion years.
Most (85\%) of this population of ejected objects originated on orbits
with perihelion distances $q < 39$ AU.  Traingles show the initial orbits
of particles ejected between $t=2$ billion and $t=3$ billion years.
Asterisks show the initial orbits of particles ejected between
$t=3$ billion and $t=4$ billion years.

We were surprised to find that the objects ejected from our simulation
between $t=3$ billion years and $t=4$ billion years
came exclusively from high initial inclinations
($i > 15^{\circ}$).  Objects in our first batch encountered Neptune at
a rate roughly proportional to $t^{-1}$ for the first 1 billion years,
in agreement with \citet{holm93} and \citet{dunc95}.
However, during the last billion years of the simulation, not one particle
from low inclinations ($i < 15^{\circ}$)
encountered Neptune, while 14 particles from higher inclinations
($15^{\circ} < i < 33^{\circ}$) encountered Neptune during this interval.
We would expect $\sim 10$ encounters for the low inclination group
during that period if the rate remained proportional to $t^{-1}$.

Naturally, the overlapping secular resonances that create
the large region of instability at low inclinations throughout the
classical Kuiper Belt affect the rate of Neptune encounters for
low-inclination objects.  These secular resonances
remove low-inclination objects early on, so relatively few remain at
$t > 3$ billion years to encounter Neptune; in the region $q < 39$ AU
almost twice as many high inclination ($15^{\circ} < i < 33^{\circ}$) objects
survive for 3 billion years as low inclination objects ($i < 15^{\circ}$).
However, this selective removal of low inclination objects does not
explain why not a single low inclination object 
in our simulation encountered Neptune from $t=3$ billion to $t=4$ billion years.
Enough low inclination objects remain at $t=4$ billion years to
provide a significant flux of Neptune encounters.

But the secular resonances do not act alone.
MMRs can protect the low-inclination objects from
destruction at the hands of the overlapping secular resonances.
The secular resonances may weed out all but the MMR-protected
low-inclination objects in the first 3 billion years, so after that time,
the only available unprotected objects for them to eject have high inclinations.

Figure~\ref{fig:simulatedae} shows that at $a < 44$ AU,
all but a few particles with $q < 39$ AU that survive at low inclinations
for 4 billion years (Figure~\ref{fig:simulatedae}) inhabit
MMRs.  High inclination particles in this
zone are spread more evenly; some high inclination particles lie in
MMRs, but many solid squares fall between the MMRs.
Figure~\ref{fig:realae} shows that the real KBOs mimic the behavior
of the test particles which survive 4 billion years;
12 out of the 15 low inclination ($i < 15^{\circ}$) KBOs at
$a < 44$ AU, $q < 39$ AU with multiple-opposition orbits
have observed semimajor axes and eccentricities within or near
the numerically calculated boundaries of MMRs.
The observed high-inclination KBOs are spread more
evenly in semimajor axis.


Our simulation suggests that todays short-period comets, Centaurs,
and scattered KBOs originate in the high-inclination KBO population.  
Therefore, short-period comets, Centaurs, and scattered KBOs should
have a color distribution like the high-inclination KBOs when they
leave the Kuiper belt.  \citet{truj02} found that low inclination
KBOs are uniformly red ($B-R \approx$ 1.5--2.2 for $i < 12^{\circ}$),
while high-inclination KBOs range from red to blue ($B-R \approx$ 1.0--2.0
for $i > 12^{\circ}$).  

\citet{jewi02} compared the colors of KBOs, Centaurs,
comet nuclei, and candidate dead comets, and found that the
Centaurs in his study had a wide range of colors.  This range
appears to be consistent with the color range of the high-inclination
KBOs in \citep{truj02}, in agreement with our dynamical result.
More color measurements of Centaurs may determine whether
the color distribution of centaurs matches the color distribution
of the high-inclination KBOs in detail.
Jewitt's samples of comet nuclei and candidate dead comets
had no objects as red as the red high-inclination
KBOs; our study offers no solution to the problem of this missing
ultrared matter.

\section{Conclusions}

Dynamical erosion is not the only long-term effect in the classical
Kuiper Belt; collisions may shape the distributions of KBO orbital
elements \citep{ster97, keny98, keny99, durd00}.  But as the collision
rate and the mass of the Kuiper Belt decreased with time, the dynamical
effects shown by our integrations must have begun to dominate the
shaping of the classical Kuiper Belt.
We found that interactions with the massive planets
preferentially deplete low-inclination objects in the inner half
of the classical Kuiper Belt, raising the mean inclination of
the population of surviving objects.  However, the outer half of the
classical Kuiper Belt ($a > 44$ AU, $q < 39 AU$) could have
retained its ancient inclination distribution.

We also found that objects ejected from the classical Kuiper Belt during
the last 1 billion years of our integration primarily come from
initial orbits at high inclinations and eccentricities.
This effect may be caused indirectly by $\nu_{17}$ and $\nu_{18}$,
which remove many low-inclination particles at early times and weed out those 
unprotected by MMRs.  This finding implies that Centaurs, short-period
comets and scattered KBOs should have an initial color distribution like
that of the high-inclination KBOs.  Measuring of the colors of
more Centaurs could test this hypothesis.

Perhaps the high-inclination KBOs separated from the primordial
KBOs objects during a special event early in the lifetime of the solar
system.  If today's active comets arise mainly from 
high-inclination KBOs, we have this event to thank for their inner
solar system company.

\acknowledgements
 
Thanks to Edo Berger, John-Philippe Berger, John Cartwright, Roy Gal,
Richard Ellis, Brian Jacoby, Ken Jucks, David Kaplan, Robert Kirshner,
David Latham, Lori Lubin, Brian Mason, Mark Metzger, Rafael Milan-Gabet,
Irene Porro, Nick Scoville, Patrick Shopbell, John Sievers, Wesley Traub,
Pat Udomprasert, and Bill Wyatt for donating CPU time for this project.
Thanks to Hal Levison, Scott Kenyon, and Ed Thommes for helpful discussions.

This work was performed in part under contract with the Jet Propulsion 
Laboratory (JPL) through the Michelson Fellowship program funded by 
NASA as an element of the Planet Finder Program.  JPL is managed for 
NASA by the California Institute of Technology

\end{document}